\documentclass[aps,pra,twocolumn,floatfix,amsmath,showpacs,superscriptaddress]{revtex4}
\usepackage{braket}
\usepackage{CJK}
\usepackage{multirow}
\usepackage{graphicx}
\usepackage{tabularx}
\usepackage{booktabs}
\usepackage{epsfig,graphicx}
\usepackage{flafter}
\usepackage{amsmath}
\usepackage{color}

\usepackage[colorlinks,urlcolor=blue,citecolor=blue,linkcolor=blue]{hyperref}

\begin{document}
\begin{CJK*}{UTF8}{gbsn} 
\title{The three-body parameter for Efimov states in lithium-6}
\author{Bo Huang(黄博)}
\affiliation{Institut f\"ur Experimentalphysik, Universit\"at
 Innsbruck, 
 6020 Innsbruck, Austria}
\author{Kenneth M. O'Hara}
\affiliation{Department of Physics, Pennsylvania State University,
University Park, Pennsylvania 16802-6300, USA}
\affiliation{Institut f\"ur Quantenoptik und Quanteninformation (IQOQI),
 \"Osterreichische Akademie der Wissenschaften, 6020 Innsbruck,
 Austria}
\author{Rudolf Grimm}
\affiliation{Institut f\"ur Experimentalphysik, Universit\"at
 Innsbruck, 
 6020 Innsbruck, Austria}
\affiliation{Institut f\"ur Quantenoptik und Quanteninformation (IQOQI),
 \"Osterreichische Akademie der Wissenschaften, 6020 Innsbruck,
 Austria}

\author{Jeremy M. Hutson}
\affiliation{Joint Quantum Centre (JQC) Durham/Newcastle, Department of
Chemistry, Durham University, South Road, Durham, DH1~3LE, United Kingdom}

\author{Dmitry S. Petrov}
\affiliation{Universit\'{e} Paris-Sud, CNRS, LPTMS, UMR8626, Orsay F-91405, France}

\date{\today}
\begin{abstract}
We present a state-of-the-art reanalysis of experimental results on Efimov
resonances in the three-fermion system of $^6$Li. We discuss different
definitions of the 3-body parameter (3BP) for Efimov states, and adopt a
definition that excludes effects due to deviations from universal scaling for
low-lying states. We develop a finite-temperature model for the case of three
distinguishable fermions and apply it to the excited-state Efimov resonance to
obtain the most accurate determination to date of the 3BP in an atomic
three-body system. Our analysis of ground-state Efimov resonances in the same
system yields values for the three-body parameter that are consistent with the
excited-state result. Recent work has suggested that the reduced 3BP for atomic systems
is a near-universal quantity, almost independent of the particular atom
involved. However, the value of the 3BP obtained for $^6$Li is significantly
($\sim 20$\%) different from that previously obtained from the excited-state
resonance in Cs. The difference between these values poses a challenge for
theory.
\end{abstract}
\maketitle
\end{CJK*}

\section{Introduction}

Ultracold atomic gases with resonant interactions provide experimental model
systems to explore the universal physics of few-body quantum states
\cite{Braaten2006uif, Ferlaino2010fyo}. Efimov states, which are weakly bound
three-body quantum states in systems of resonantly interacting particles, are a
paradigm of this field. Efimov \cite{Efimov1970ela} showed that, when two
bosons interact with an infinite scattering length, the corresponding
three-particle system has an infinite number of three-body states just below
threshold. For zero-range interactions, each successive Efimov state is larger
than the previous one by a discrete length scaling factor, the `Efimov period',
which is 22.7 for a system of three identical bosons \cite{note:efactor} but
can be widely different for other systems \cite{Dincao2006eto}. We refer to
this universal scaling behavior as Efimov universality.

The interactions between pairs of ultracold atoms may be varied by tuning an
applied magnetic field in the vicinity of a zero-energy Feshbach resonance
\cite{Chin2010fri}. The scattering length has a pole at resonance,
corresponding to a 2-body bound state exactly at threshold. Signatures of
Efimov states were first observed in an ultracold gas of cesium atoms
\cite{Kraemer2006efe}, and have since been found in many other systems,
including other bosonic gases \cite{Zaccanti2009ooa, Pollack2009uit,
Gross2009oou, Gross2010nsi, Wild2012mot, Roy2013tot}, three-component fermionic
spin mixtures \cite{Ottenstein2008cso, Huckans2009tbr, Williams2009efa,
Nakajima2010nea}, and mixtures of atomic species \cite{Barontini2009ooh,
Bloom2013tou, Tung2014oog, Pires2014ooe}. Moreover, extensions of the Efimov
scenario to universal states of larger clusters \cite{Hammer2007upo, vonStecher2009sou, vonStecher2010wbc} have been demonstrated in
experiments \cite{Ferlaino2009efu, Pollack2009uit, Zenesini2013rfb},
highlighting the general nature of universal few-body physics.

In addition to their discrete scaling property, Efimov states are characterized
by a {\it three-body parameter} (3BP), which determines the position of the
entire ladder of states. In the realm of nuclear systems, the 3BP is a
non-universal quantity \cite{Braaten2006uif}, determined by details of the
short-range interaction. However, in atomic systems it has been found
experimentally \cite{Berninger2011uot, Roy2013tot} that the 3BP is nearly
constant when expressed in terms of the van der Waals length $r_{\rm vdW}$
\cite{Chin2010fri}, which quantifies the dispersion interaction between two
neutral atoms. We refer to this feature of Efimov physics as van der Waals
universality of the 3BP, and it has been the subject of a number of theoretical
investigations \cite{Chin2011uso, Wang2012oot, Schmidt2012epb, Sorensen2012epa, Naidon2014moa,
Wang2014uvd}.

Three-body recombination resonances occur when Efimov states cross the
three-atom threshold as a function of magnetic field (and hence of scattering
length) \cite{Esry1999rot, Braaten2001tbr, Ferlaino2011eri}. Recombination
resonances due to Efimov ground states provide the most prominent observables
in Efimov physics. Many experiments have focused on such features, including
some that determined the 3BP \cite{Berninger2011uot, Wild2012mot, Roy2013tot}.
In real atomic systems, however, finite-range corrections may significantly
affect universal scaling, particularly for ratios involving the Efimov ground
state \cite{Thogersen2008nbe, Platter2009rct, Naidon:2011, Schmidt2012epb}.
However, such corrections decrease substantially for higher Efimov states and
are already very small for the first excited state. Excited-state resonances
are therefore particularly interesting for precise measurements of the 3BP.

Excited-state Efimov resonances occur at very large scattering lengths. They
require extremely low temperatures for experimental observation, since the
recombination peaks are less well defined when the de Broglie wavelengths are
shorter than the scattering lengths \cite{Dincao2004lou, Kraemer2006efe,
Rem2013lot}. Excited-state resonances have therefore been observed in only a
very few experiments, carried out with $^6$Li \cite{Williams2009efa}, with
$^{133}$Cs \cite{Huang2014oot}, and with mixtures of $^6$Li and $^{133}$Cs
\cite{Tung2014oog, Pires2014ooe}. Quantitative understanding of these
resonances requires both very precise knowledge of the two-body scattering
properties and an accurate theoretical description of finite-temperature
effects. Ref.~\cite{Huang2014oot} analyzed the excited-state Efimov resonance
in cesium, using a highly accurate model of the two-body scattering
\cite{Berninger2013frw} and a theoretical finite-temperature approach recently
developed in Ref.~\cite{Rem2013lot}. This study provided the most precise
measurement of the Efimov period so far.

In this Article, we re-analyze previous experimental results on the
excited-state Efimov resonance in $^6$Li observed in
Ref.~\cite{Williams2009efa} and on the ground-state Efimov resonances observed
in Refs.~\cite{Ottenstein2008cso} and \cite{Huckans2009tbr}. In
Sec.~\ref{sec:3BP}, we discuss different definitions of the 3BP and how they
are affected by deviations from ideal Efimov behavior. We adopt a definition
that excludes effects due to deviations from universal scaling for low-lying
states. In Sec.~\ref{sec:threefermion}, we summarize the main properties of the
three-fermion system. In Sec.~\ref{sec:model}, we develop a new
finite-temperature approach, which generalizes the theory introduced for the
three-boson case in Ref.~\cite{Rem2013lot} to the case of three distinguishable
fermions. In Sec.~\ref{sec:excited}, we present a refined analysis of the
excited-state resonance observed in Ref.~\cite{Williams2009efa}. This gives a
high-precision value for the 3BP in $^6$Li, which deviates significantly from
those found in other atomic systems. In Sec.~\ref{sec:ground}, we re-analyze
previous results on the ground-state Efimov resonance from
Ref.~\cite{Ottenstein2008cso} and investigate the possible influence of
finite-range effects. In Sec.~\ref{sec:discuss}, we discuss our findings in the
context of other experiments in the field. The value of the 3BP found for
$^6$Li is not well explained by current theories and presents a challenge for
future theoretical work.

\section{Three-body parameter}
\label{sec:3BP}

For three identical bosons, ideal Efimov scaling leads to the simple relation
\begin{equation}
\kappa^{(n+1)} = \kappa^{(n)} / 22.7 \,
\label{eq:kappa_universal}
\end{equation}
between the wavenumbers $\kappa^{(n)}$ that characterize the energies
$E^{(n)}_{\rm res} = - (\hbar \kappa^{(n)})^2/m$ of successive Efimov states in
the resonant limit $a \rightarrow \pm\infty$. Here $m$ is the atomic mass and
$n$ is an integer quantum number. The corresponding relation between the
the scattering lengths at the positions of successive recombination resonances is
\begin{equation}
a_-^{(n+1)} = 22.7 \times a_-^{(n)} \, .
\label{eq:a_universal}
\end{equation}
The universal relation
\begin{equation} a_-^{(n)} = -1.508 / \kappa^{(n)} \,
\label{eq:aminus}
\end{equation}
connects a resonance position with the corresponding bound-state wavenumber. In
the ideal case, knowledge of any of the above quantities $\kappa^{(n)}$ or
$a_-^{(n)}$ fixes the infinite series and thus provides a proper representation
of the 3BP.

In a real system, where the interaction has a finite range, the Efimov spectrum
is bounded from below. We refer to the lowest state as the Efimov ground state
with $n=0$ and to the corresponding resonance at $a_-^{(0)}$ as the
ground-state Efimov resonance. Eqs.~(\ref{eq:kappa_universal}) and (\ref{eq:a_universal})
then represent approximations, subject to finite-range effects.

One way to understand the Efimov effect is through a treatment in
hyperspherical coordinates. Efimov states may be viewed as supported by an
effective adiabatic potential that is a function of the hyperradius $R$. For a
zero-range two-body potential with large scattering length $a$, this potential is
attractive and proportional to $R^{-2}$ for $R\lesssim|a|$ \cite{Macek:1986} and
supports an infinite number of bound states as $a\rightarrow\pm\infty$. For
potential curves with long-range van der Waals tails, however, Wang {\em et
al.} \cite{Wang2012oot} have shown that the effective adiabatic potential
reaches a minimum and then rises to a wall or barrier near $R=2r_{\rm vdW}$. The position of the minimum and wall depend to some extent on the details of
the two-body potential and the number of bound states it supports, but become
near-universal as the number of 2-body bound states increases. The presence of
the minimum and wall have two principal effects on the physics. First, the
deviation of the effective potential from $R^{-2}$ behavior results in
deviations from ideal Efimov scaling for the lowest-lying states. Secondly, the
boundary condition provided by the wall defines the position of the entire
ladder of Efimov states, and its nearly universal position is responsible for
the near-universality of the 3BP. However, it should be noted that the wall
itself is a product of physics around $2r_{\rm vdW}$, so that variations in the
physics in this region can produce deviations from universality of the 3BP even
in the limit $a\rightarrow\pm\infty$.

Theoretical investigations \cite{Thogersen2008nbe, Platter2009rct, Wang2012oot,
Schmidt2012epb} have shown that the Efimov ground state may be subject to
considerable modifications. For $n=0$ this may change the factor $22.7$ in
Eqs.~(\ref{eq:a_universal}) - (\ref{eq:aminus}) by up to 25\%. The relation
(\ref{eq:aminus}) is subject to similar modifications \cite{Schmidt2012epb,
Wang2012oot}. The recent experiment on the excited-state resonance in Cs
\cite{Huang2014oot} and a related theoretical investigation \cite{Wang2014uvd}
also hint at deviations from the ideal scaling.

The deviations from universal scaling for low-lying Efimov states raise the
question of the best representation of the 3BP. Definitions based on the limit
$n \rightarrow \infty$ remove effects of this type from the 3BP. Accordingly,
we adopt the definition \cite{Braaten2006uif}
\begin{equation}
\kappa_* = \lim_{n \to \infty} \left( 22.7^n \kappa^{(n)} \right) \, ,
\label{eq:idealkappa}
\end{equation}
and by analogy
\begin{equation}
a_-^* = \lim_{n \to \infty} \frac{a_-^{(n)}}{22.7^n}\, .
\label{eq:ideala}
\end{equation}

The position of the ground-state Efimov resonance, $a_-^{(0)}$, is commonly
used as a 3BP. However, it gives a somewhat crude approximation to $a_-^*$, and
in some cases may deviate from it by as much as 25\%. The quantities
$a_-^{(1)}/22.7$ and $22.7\kappa^{(1)}$, obtained from the excited-state
resonance, provide much better approximations to $a_-^*$ and $\kappa_*$, with
corrections of only about 1\% due to deviations from universal scaling
\cite{Schmidt2012epb}; these corrections are comparable to the other
uncertainties in current experiments.

Efimov states are also characterized by a decay parameter $\eta_*$
\cite{Braaten2006uif}, which describes their decay to lower-lying atom-dimer
combinations. This parameter is usually considered to be a constant for a
particular Efimov state, but may vary if the available product states change
significantly. The resulting field dependence may be important when
interpreting measurements that extend over wide ranges of field
\cite{Wenz2009uti}.

\section{Efimov states in a three-component Fermi gas}
\label{sec:threefermion}

\subsection{Three-fermion system}

Efimov states in a three-component gas of fermions \cite{Braaten2009tbr}
exhibit the same discrete scaling behavior as in the three-boson case, provided
that all three scattering lengths involved are large ($|a_{12}|, |a_{13}|,
|a_{23}| \gg r_{\rm vdW}$). In particular, if the masses of the three
components are equal, the Efimov period is given by the same discrete scaling
factor of $22.7$ \cite{note:efactor}. The special case of three equal
scattering lengths ($a_{12} = a_{13} = a_{23}$) is formally equivalent to the
situation for three identical bosons.

A gas of $^6$Li atoms prepared in a mixture of the lowest three spin states
allows a realization of large scattering lengths by Feshbach tuning
\cite{Chin2010fri}. However, the applied magnetic field offers only one degree
of freedom for tuning, thus limiting the experimentally accessible combinations
of scattering lengths. Arbitrary combinations and, in particular, the situation
of three equal scattering lengths thus remain hypothetical cases, but universal
theory allows them to be linked to the combinations that exist in real systems.

In real experiments on a three-fermion system, Efimov resonances appear at
certain combinations of large scattering lengths $a_{12}, a_{13}, a_{23}$,
where typically $a_{12} \neq a_{13} \neq a_{23}$. A generalization of the
Skorniakov--Ter-Martirosian (STM) equations \cite{Braaten2009tbr} can be
employed to determine the 3BP from these generally unequal values. In the
wavenumber representation, $\kappa_*$ then refers to the hypothetical case of
three infinite scattering lengths, while $a_-^*$ refers to a
hypothetical system with three equal scattering lengths.

The STM approach is based on the zero-range approximation and therefore does
not take account of finite-range corrections, which are significant at 
relatively small scattering lengths. It can thus be expected to provide an excellent approximation for excited Efimov states ($n \ge 1$), but it may be subject to significant
corrections if applied to the Efimov ground state ($n=0$).

\subsection{Three-body recombination}

In a three-component Fermi gas, the dominant contribution to three-body losses
results from triples of three non-identical particles. All other combinations
involve pairs of identical fermions, which leads to a strong Pauli suppression
of losses at ultralow temperatures \cite{Esry2001tlf}.

Three-body losses can be modeled by the simple rate equation
\begin{equation}
\frac{d}{dt} n_i = - L_3 n_1 n_2 n_3 \, ,
\end{equation}
where the $n_i$ represent the number densities of the three different spin
states. After a spatial integration of losses over the density profile of the
trapped cloud, the loss rate coefficient $L_3$ can be experimentally determined
by fitting the time-dependent decay of the total atom numbers
\cite{Ottenstein2008cso, Huckans2009tbr, Williams2009efa}. Efimov states show
up as distinct loss resonances \cite{Ferlaino2011eri} when they couple to the
three-atom threshold.

\subsection{Lithium-6}

The situation of a three-component Fermi gas of $^6$Li is unique because of
overlapping Feshbach resonances in all three combinations of the lowest three
spin states together with large negative background scattering lengths. The
two-body scattering properties are known to an extraordinarily high level of
precision thanks to the characterization in Ref.~\cite{Zurn2013pco}, which
significantly improved the conversion from magnetic field to scattering lengths
compared to previous work \cite{Bartenstein2005pdo}.

In the resonance region between 832 and 900\,G, all three scattering lengths
are very large and negative, with absolute values of a few thousand times the
Bohr radius $a_0$ that vastly exceed $r_{\rm vdW} = 31.26\,a_0$. In this
extreme regime, an excited Efimov state exists \cite{Williams2009efa}. This
trimer state crosses the three-atom threshold near 900\,G and leads to a strong
enhancement of three-body recombination. The corresponding Efimov ground state
exists over a much wider range of magnetic fields, but it does not cross
threshold at currently accessible magnetic fields 
and thus does not lead to an observable recombination resonance.

In the magnetic-field region below the zero crossings of the Feshbach
resonances, the three scattering lengths are moderately large and negative, so
that an Efimov ground state exists. This state crosses the three-atom threshold
near 130\,G and near 500\,G \cite{Ottenstein2008cso, Huckans2009tbr}, leading
to two observable Efimov resonances. In this low-field region, the scattering
lengths never reach large enough values for an excited Efimov state to exist.

\subsection{The effective range}

One way to quantify the finite (non-zero) range of an atomic interaction is
through the effective range \cite{Bethe:1949,Hinckelmann:1971}, which
characterizes the leading term in the energy-dependence of the scattering
length. The effective range behaves very differently in the vicinity of
Feshbach resonances of different types \cite{Blackley:eff-range:2014}. For a
resonance that is strongly entrance-channel-dominated \cite{Chin2010fri}, the
effective range takes a small and fairly constant value close to $2.8r_{\rm vdW}$ at fields near the resonance pole \cite{Gao:QDT:1998}. By contrast, for
resonances that are closed-channel-dominated, the effective range is much
larger and varies very fast with magnetic field \cite{Blackley:eff-range:2014}.
The Feshbach resonances used in the present work for $^6$Li are all strongly
entrance-channel-dominated \cite{Chin2010fri}, so that deviations from Efimov
scaling due to finite-range effects are expected to be relatively small in
comparison to some of the other atomic systems that have been studied.

\section{Finite-temperature theoretical approach} \label{sec:model}

A convenient way of modeling three-body losses in Efimovian systems is provided
by the $S$-matrix formalism based on Efimov's radial law \cite{Efimov1979lep}, 
which is elaborated in Refs.~\cite{Braaten2006uif,Braaten2008tbr,Rem2013lot} for the
case of three identical bosons. Its generalization to three distinguishable
atoms with different scattering lengths is straightforward and we will present
only a brief derivation. This is a zero-range theory for which $\kappa_*$ and $\eta_*$ are external parameters.

First, one introduces three-atom scattering channels describing the motion of free atoms at large distances. 
By contrast, all atom-dimer channels are substituted by the
single Efimov channel defined in the scaling region $r_{\rm vdW}\ll R\ll {\rm
min}\{1/k,|a_{12}|,|a_{23}|,|a_{13}|\}$, where $k=\sqrt{mE}/\hbar$, $E$ is the
energy in the center of mass reference frame, $R$ is the hyperradius, and we
consider the case of negative scattering lengths. The reason for this
substitution is that when $r_{\rm vdW}\ll {\rm
min}\{1/k,|a_{12}|,|a_{23}|,|a_{13}|\}$ this channel becomes essentially the
only one that can conduct three atoms from large distances to the recombination
region of size $\sim r_{\rm vdW}$.

One can think of this short-distance channel and the long-distance three-atom
channels as being fused together at intermediate distances where the
transmission, reflection, and mixing of the channels takes place. We can then
introduce a unitary matrix $s_{ij}$, which defines the amplitude of the
outgoing wave in channel $j$ if the incoming wave is injected in
channel $i$. The terms ``incoming'' and ``outgoing'' are defined with respect
to the fusion region. In particular, the incoming Efimov wave $R^{-2+is_0}$
actually propagates towards larger distances and $R^{-2-is_0}$ describes the
outgoing one. Here $s_0 \approx 1.00624$ is a constant and the
ideal Efimov period of 22.7 is $e^{\pi/s_0}$.

The simple fact that the matrix $s_{ij}$ is unitary turns out to be very useful
in describing the scaling properties of Efimovian systems
\cite{Braaten2006uif}. We point out that $s_{ij}$ does not depend on the
3BP $\kappa_*$ or the decay parameter $\eta_*$. These
quantities come into play when one fixes the relative phase and amplitude of
the incoming and outgoing Efimov waves,
\begin{equation}\label{ThreeBodyCond}
R^2\Psi\,{\propto}\,(R/R_0)^{is_0}{-}e^{2\eta_*}(R/R_0)^{-is_0},
\end{equation}
where $R_0$ is a three-body length related to $\kappa_*$ by
\begin{equation}
(\kappa_* R_0/2)^{2is_0}=-\Gamma(is_0)/\Gamma(-is_0)
\end{equation}
and $\Gamma$ is the gamma function.
One can imagine that Efimov waves are reflected at small hyperradii by a lossy
mirror with reflection/loss properties given by Eq.~(\ref{ThreeBodyCond}). The
three-body problem is then analogous to a Fabry-Perot interferometer with the
other mirror quantified by the matrix $s_{ij}$. This picture gives a convenient
way of understanding and describing three-body loss peaks as resonances of the
Fabry-Perot cavity. In particular, if we denote the Efimov channel by subscript
1, the loss probability for a given incoming channel $i\neq 1$ is
\cite{Braaten2008tbr}
\begin{equation}
P_i=\frac{(1-e^{-4\eta_*})|s_{i1}|^2}{|1+(kR_0)^{-2is_0}e^{-2\eta_*}s_{11}|^2},
\end{equation}
where the denominator accounts for multiple reflections ``inside'' the
resonator. The total loss rate constant for three distinguishable fermions is
obtained by using unitarity ($\sum_{i=1}^\infty|s_{1i}|^2\,{=}\,1$) and
averaging over the Boltzmann distribution,
\begin{eqnarray}\label{L3result}
L_3&=&\frac{24\sqrt{3}\pi^2\hbar (1-e^{-4\eta_*})}{mk_{\rm th}^{6}} \nonumber\\
&\times& \int_0^\infty\frac{(1-|s_{11}|^2)e^{-k^2/k_{\rm th}^2}}{|1+(kR_0)^{-2is_0}e^{-2\eta_*}s_{11}|^2}k\,dk,
\end{eqnarray}
where $k_{\rm th}=\sqrt{mk_{\rm B}T}/\hbar$. Equation~(\ref{L3result}) differs
from the bosonic result of Ref.~\cite{Rem2013lot} only by the factor $1/3$,
which is due to the bosonic bunching effect and different ways of counting
triples in the two cases. A more profound change is hidden in the quantity
$s_{11}$, which, in contrast to the case of identical bosons, now depends on
three dimensionless numbers $ka_{12}$, $ka_{23}$, and $ka_{13}$. 

In order to determine $s_{11}$ we look for the three-body wave function that
behaves as $A (kR)^{-2+is_0}+ B (kR)^{-2-is_0}$ in the scaling region and contains
only outgoing waves at large distances. By definition, $s_{11}=B/A$. We solve
this problem by using the STM equations in a
very close analogy to the bosonic case (see Supplemental Material of
\cite{Rem2013lot}). For distinguishable atoms with generally different
scattering lengths we end up with three coupled STM equations (see
Ref.~\cite{PetrovLesHouches2010} for details of the method). 

In practice, we use the known dependence of $a_{ij}$ on $B$ \cite{Zurn2013pco}
and tabulate $s_{11}$ as a function of $k$ and $B$. This then allows fast
integration of Eq.~(\ref{L3result}) for any desired values of $T$, $\kappa_*$,
and $\eta_*$.

\section{Excited-state Efimov resonance}
\label{sec:excited}

\begin{figure}
\includegraphics[width=8.6cm]{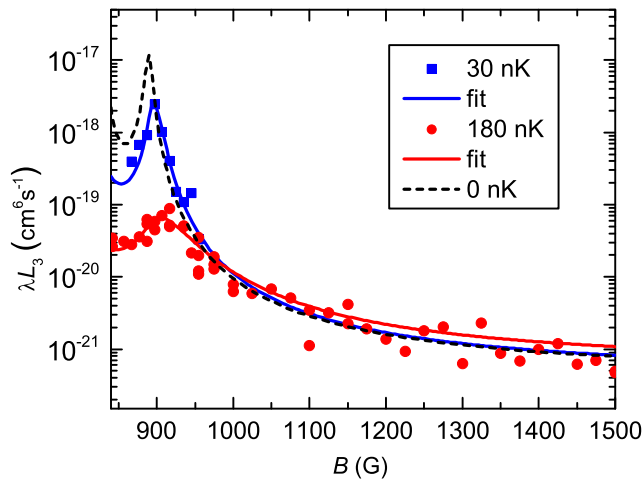}
\caption{(Color online). Finite-temperature fits to the excited-state Efimov
resonance. The experimental results obtained for $L_3$ in
Ref.~\cite{Williams2009efa} for two different temperatures are plotted as
filled blue squares (set A, 30\,nK) and filled red circles (set B, 180\,nK).
The amplitude scaling parameter $\lambda$ is of order 1, see text. The
corresponding solid lines are the fixed-temperature fits to both data sets,
carried out on a linear scale (see first and fifth row in Table~\ref{tab.2nd}).
The black dashed curve is calculated for the zero-temperature limit using the
parameters from the fixed-temperature fit to the 30~nK set.} \label{fig.2nd}
\end{figure}

In Ref.~\cite{Williams2009efa}, the excited-state Efimov resonance was observed
in the high-field region of $^6$Li. In Figure~\ref{fig.2nd} we show the
experimental results for the three-body loss coefficient $L_3$ as a function of
the magnetic field, measured for two different temperatures of about 30\,nK
(set A) and 180\,nK (set B). In this section we reanalyze these results, taking
account of finite-temperature effects using the theory described in
Sec.~\ref{sec:model}, in order to obtain a refined estimate of the 3BP for
$^6$Li.

The two free parameters in the temperature-dependent theory of
Sec.~\ref{sec:model} are the 3BP $\kappa_*$ and the decay parameter $\eta_*$.
In addition, experimental uncertainties in the number density calibration may
considerably affect the amplitude of the observed losses. Such uncertainties
may result from the atom number calibration, from the limited knowledge of the
trap frequencies, and from errors in the temperature measurements. It is
therefore useful to introduce an additional scaling parameter $\lambda$ for the
amplitude of the observed losses \cite{Huang2014oot}. Under realistic
experimental conditions, variations of up to a factor of two from the ideal
value $\lambda = 1$ are plausible.

To analyze the data we follow several different strategies, similar to those
applied to the three-boson case of cesium \cite{Huang2014oot}. First, we fix
the temperature $T$ to the measured values $T_{\rm meas} = 30$\,nK (set A) and
180\,nK (set B), and we perform a fit with $\kappa_*$, $\eta_*$, and $\lambda$
as the free parameters. Alternatively, we allow for a variable temperature $T$,
and instead we fix $\lambda = (T/T_{\rm meas})^{-3}$ \cite{note:unstable} to take
account of the resulting change in the volume of the harmonically trapped gas.
Moreover, we fit the data sets A and B on either a linear or a logarithmic
scale, which puts different weights on the different regions. In this way, we
obtain four different fits for each data set. We note that the experimental results of Ref.~\cite{Williams2009efa} indicated that the effect of heating during the decay of the trapped sample remained very small, so that this effect can safely be neglected in our fit analysis.

\begin{table}
\begin{ruledtabular}
\begin{tabular}{lrrrr}
Set       & $T$ (nK) 	& $\kappa_* a_0$ & $\eta_*$ & $\lambda$ \\
\hline
A         & 30$^a$  & 0.006808(36) & 0.032(5)  & 0.546(27)  \\
A log     & 30$^a$  & 0.006744(91) & 0.048(15) & 0.498(107)  \\
A         & 35(5)   & 0.006774(39) & 0.029(5)  & 0.644$^T$  \\
A log     & 36(2)   & 0.006689(97) & 0.042(14) & 0.593$^T$  \\
\hline
B         & 180$^a$ & 0.006839(80) & 0.088(15) & 0.258(16)  \\
B log     & 180$^a$ & 0.006665(130)& 0.067(16) & 0.270(49)  \\
B         & 237(5)  & 0.006736(84) & 0.072(15) & 0.438$^T$  \\
B log     & 218(10) & 0.006624(118)& 0.034(8)  & 0.562$^T$  \\
\end{tabular}
\end{ruledtabular}
\caption{Results of fits for the excited-state Efimov resonance, obtained from the two sets of measurements presented in Fig.~\ref{fig.2nd}.
The fits using a logarithmic $L_3$ scale are indicated with `log' in the column
`Set'. The superscript $^a$ means that corresponding parameter is kept fixed.
The superscript $^T$ indicates that the corresponding parameter is calculated
from the fitted values for $T$.} \label{tab.2nd}
\end{table}

\begin{figure}
\includegraphics[width=8.6cm] {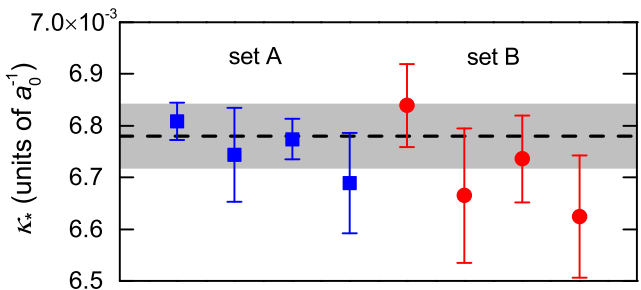}
\caption{(Color online). Fitted values for $\kappa_*$ corresponding to the
third column in Table~\ref{tab.2nd}. The dashed line indicates the final result
$\kappa_*  = 0.00678(6)\,a_0^{-1}$, as obtained from a weighted average of the
four data points of the low-temperature data set A (filled blue squares), and
the gray-shaded region shows the corresponding uncertainty. The
high-temperature data set B (filled red circles) is not used for deriving the
final value, but within the uncertainties the values are fully consistent with
the result from data set A.} \label{fig.kappas}
\end{figure}

Table \ref{tab.2nd} summarizes the results of our fits for both data sets, and
Fig.~\ref{fig.kappas} shows the values obtained for the 3BP (from the third
column of the Table). The comparison between the four different fits for each
data set provides information on the robustness of the fits and possible
systematic effects beyond simple statistical uncertainties. In our results from
the low-temperature set A, the errors on $\kappa_*$ from individual fits range
between 0.5\% (for linear fits) and 1.5\% (for logarithmic fits). Within the
error bars no significant systematic deviations appear between the central
values obtained from the different fits, which shows that the errors are
consistent with purely statistical uncertainties. From the low-temperature data
set~A, by calculating a weighted average \cite{note:weights} over all four
fitted values, we obtain the final value $\kappa_*  = 0.00678(6)\,a_0^{-1}$,
where the uncertainty includes both the weighted errors of the four individual
fits and the standard deviation of the four slightly different values. The
result for $\kappa_*$ and the error are shown by the dashed horizontal line and
the gray-shaded region in Fig.~\ref{fig.kappas}. Note that all the statistical
uncertainties specified in this work correspond to one standard deviation.

The higher-temperature data set B yields similar results, but with somewhat
larger uncertainties. Again, there are no systematic deviations between the
four different fit strategies applied. Here the final result for the 3BP,
$\kappa_* = 0.00674(13)\,a_0^{-1}$, is fully consistent with the result
obtained at lower temperatures, with an uncertainty about two
times larger than for set A. This confirms that temperature-induced shifts of
the resonance are properly taken into account in our theoretical approach.

The original data analysis in Ref.~\cite{Williams2009efa} yielded $\kappa_* =
0.0069(2)\,a_0^{-1}$, remarkably close to the present result but with a quoted
error about three times larger. However, the present work reveals two
systematic shifts, which in the previous work partially canceled each other.
The updated values of the scattering lengths \cite{Zurn2013pco} shift the value
of $\kappa_*$ up by about 3\%, while residual finite-temperature effects shift
the value down by about 5\% \cite{note:residualT}.

A further contribution to our error budget comes from the uncertainty in the
mapping from magnetic field to scattering length. The scattering lengths used
here were obtained from the potential curves of Ref.~\cite{Zurn2013pco}, which
were fitted to highly precise measurements of binding energies of $^6$Li$_2$ in
the resonant region, together with measurements of collision properties. The
fits have recently been extended to include binding energies for $^7$Li$_2$,
with an explicit mass-dependence of the potential curves
\cite{Julienne:Li67:2014}. In order to establish the uncertainties in the
scattering lengths at the magnetic field of the excited-state resonance, we
have repeated the fits of Ref.~\cite{Zurn2013pco} and calculated explicit
statistical uncertainties in the three scattering lengths $a_{12}$, $a_{13}$
and $a_{23}$ at 891\,G, using the procedure of Ref.\ \cite{LeRoy:1998}. The
resulting contribution to the uncertainty in $\kappa_*$ is about 0.1\%. We have
also estimated the nonstatistical uncertainties in the scattering lengths by
repeating the fits with the experimental binding energies set to the values at
the upper and lower limits of their systematic errors, producing a further
uncertainty of 0.07\%. The uncertainty of 0.1\,G in the magnetic-field
calibration of Ref.~\cite{Williams2009efa} corresponds to a further error of
0.07\%. All these error sources are thus negligibly small compared to the
fitting errors described above.

Based on the results of our fits for $\kappa_*$ and $\eta_*$, we can calculate
the recombination rate coefficient $L_3$ in the zero-temperature limit. The
resulting curve is shown as a dashed line in Fig.~\ref{fig.2nd}. The peak
occurs at 891\,G, which marks the point where the Efimov state crosses the
three-atom threshold. Here the three scattering lengths are $a_{12} =
-8671(38)\,a_0$, $a_{13} = -2866(3)\,a_0$, and $a_{23} = -5728(16)\,a_0$.

\section{Ground-state Efimov resonances}
\label{sec:ground}

References~\cite{Ottenstein2008cso, Huckans2009tbr} reported the observation of
the two ground-state Efimov resonances in the low-field region of $^6$Li near
130\,G and near 500\,G. The $L_3$ results of Ref.~\cite{Ottenstein2008cso} have
been further analyzed in Refs.~\cite{Braaten2009tbr, Wenz2009uti, Naidon:2011},
using different models within the zero-temperature approximation.
Ref.~\cite{Braaten2009tbr} treated the three different scattering lengths
within the approach of the generalized STM equations, which is exact within the
zero-range limit, while Ref.~\cite{Wenz2009uti} introduced the approximation of
an `effective scattering length'. As an important improvement,
Ref.~\cite{Wenz2009uti} introduced a magnetic-field dependence in the decay
parameter $\eta_*$, determined by the binding energies of the different target
molecular states. The latter has proved very useful to describe the different
widths of the narrower Efimov resonance near 130\,G and the wider Efimov
resonance near 500\,G. Ref.\ \cite{Naidon:2011} considered the effects of
finite-range corrections and of a two-channel model of the atom-atom
scattering.

Our new analysis of the results of Ref.~\cite{Ottenstein2008cso} is based on
the generalized STM approach in combination with the magnetic-field-dependent
decay parameter $\eta_*$. We also use the updated scattering length values from
Ref.~\cite{Zurn2013pco}, instead of the ones from
Ref.~\cite{Bartenstein2005pdo}, but this has negligible effect on the value
resulting for the 3BP in the low-field region. All our fits assume a
temperature of 215\,nK \cite{Ottenstein2008cso}, but we find that including
finite-temperature effects makes a negligible difference for the ground-state
resonances, in contrast to the excited-state case.

Our theoretical model to calculate $L_3$ from the three different scattering
lengths relies on the zero-range approximation, and is applicable only for
$|a_{12}|, |a_{13}|, |a_{23}| \gg r_{\rm vdW}$. However, at the resonance
positions of 130\,G and 500\,G, the smallest of the three scattering lengths,
$|a_{12}|$, exhibits rather small values of $\sim4\,r_{\rm vdW}$ and
$\sim3\,r_{\rm vdW}$, respectively. This makes the analysis quite vulnerable
to finite-range effects, and the extracted values for $\kappa_*$ can be
expected to provide only an approximation to the limiting case of
Eq.~(\ref{eq:idealkappa}). To deal with this nonideal situation, we introduce a
lower cutoff scattering length $a_{\rm min}$ and restrict our fit to the region
where $|a_{12}|, |a_{13}|, |a_{23}| > a_{\rm min}$. The dependence of the
resulting values for $\kappa_*$ on $a_{\rm min}$ then gives an indication of
the sensitivity to finite-range and model-dependent corrections.

\begin{figure}
\includegraphics[width=8cm] {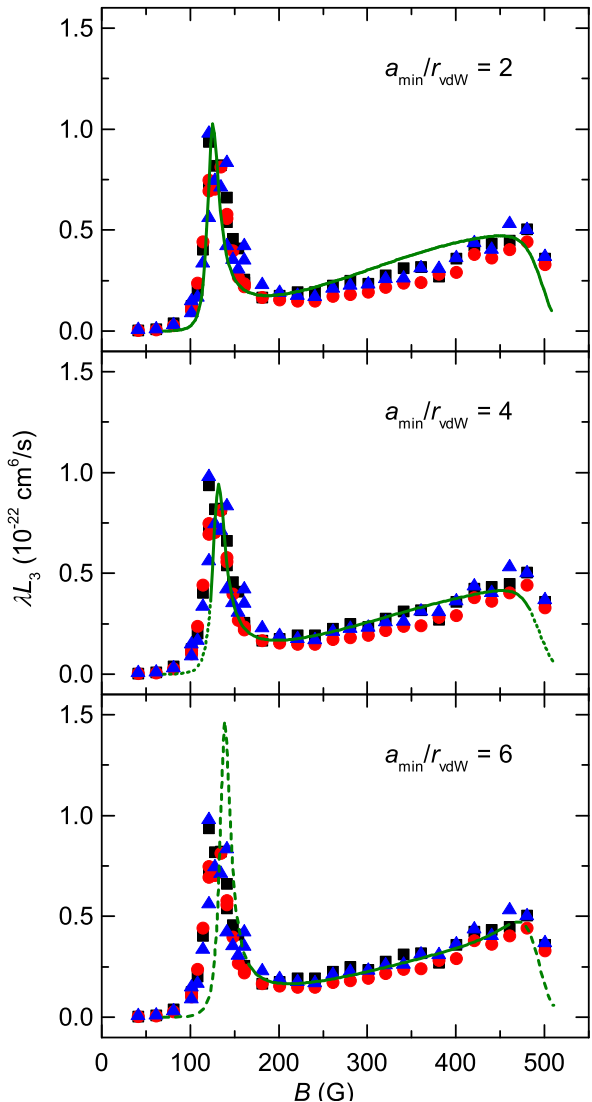}
\caption{(Color online). Fits to the ground-state Efimov resonances. All three
panels show the same experimental data on the loss rate coefficient $L_3$ from
Ref.~\cite{Ottenstein2008cso}, where the filled squares, filled circles, and
filled triangles refer to losses measured in the lowest three spin states. The
theoretical curves represent our fits to the data on a linear scale. The solid
lines indicate the region used for the fit with all three scattering lengths
being larger than the cutoff value $a_{\rm min}$. The dashed lines extrapolate
the theory to regions not used for the fit.} \label{fig.1stfit}
\end{figure}

Figure \ref{fig.1stfit} shows three different fits to the same data points,
differing in the cutoff scattering length, $a_{\rm min}/r_{\rm vdW} = 2, 4$,
and $6$. The fits are applied globally to both resonances, appearing near 130\,G and near 500\,G. The three free parameters of the fit are $\kappa_*$, the amplitude
scaling factor $\lambda$ (see Sec.~\ref{sec:excited}), and the parameter $A$
defined in Ref.~\cite{Wenz2009uti}, from which the magnetic-field-dependent
$\eta_*$ can be calculated. The lines in Fig.~\ref{fig.1stfit} represent the
behavior within the fit region (solid lines) and extrapolated beyond that
region (dashed lines). We find that the fit with $a_{\rm min}/r_{\rm vdW} = 4$
captures both resonances and the overall behavior quite well. Here we obtain
$\kappa_* = \, 0.00645(3) \, a_0^{-1}$ (linear scale) and $0.00641(3) \,
a_0^{-1}$ (logarithmic scale). For the amplitude scaling factor the fits yield
the plausible values $\lambda = 1.65(5)$ (linear) and $1.68(7)$ (logarithmic).
From the corresponding values of $A$ we obtain the values $\eta_*=0.0814(3)$
(linear) and $0.0745(3)$ (logarithmic) for the decay parameter at the
lower-field (sharper) resonance, which the fit locates at 132\,G.

In contrast to the fit with $a_{\rm min}/r_{\rm vdW} = 4$, the two other fits
in Fig.~\ref{fig.1stfit} are problematic. The fit for $a_{\rm min}/r_{\rm vdW}
= 2$ puts some weight on regions where the applicability of zero-range theory
is highly questionable, while the fit for $a_{\rm min}/r_{\rm vdW} = 6$
excludes the centers of the two resonances, which provide the most sensitive
information on the Efimov resonance positions.

\begin{figure}
\includegraphics[width=8.6cm] {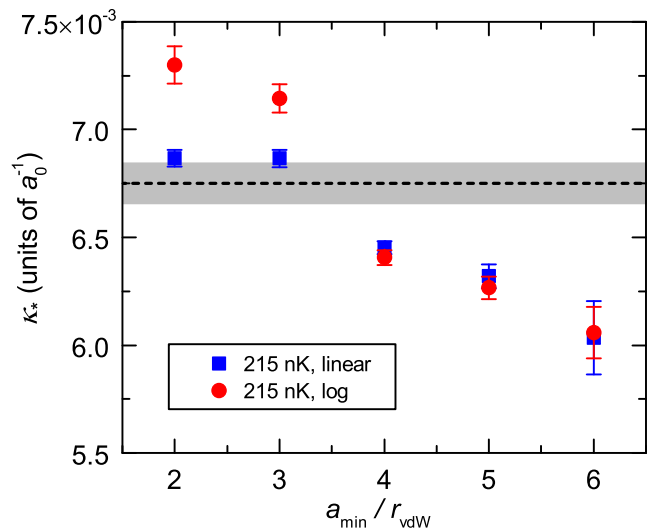}
\caption{(Color online). Dependence of the fitted values for $\kappa_*$ on the
cutoff scattering length $a_{\rm min}$ for the ground-state Efimov resonance.
The blue filled squares and red filled circles refer to fits performed with a
linear and logarithmic $L_3$ scale. The error bars represent the $1\sigma$
uncertainties from the individual fits. The horizontal dashed line marks the
value of $\kappa_*$ obtained from the excited-state Efimov resonance. The
gray-shaded region marks the corresponding error range.} \label{fig.1stBRange}
\end{figure}

Figure \ref{fig.1stBRange} shows the values for $\kappa_*$ resulting from fits
with different cutoff scattering lengths $a_{\rm min}$ in the range between $2$
and $6\,r_{\rm vdW}$. The filled blue squares represent the fit to the $L_3$
results on a linear scale. This fit puts most weight on the lower resonance,
but as $a_{\rm min}$ increases it gives more weight to the region between the
resonances, and the resulting value for $\kappa_*$ decreases by almost 10\%.
The fits to the $L_3$ data on a logarithmic scale (filled red circles) show a
similar behavior with a trend towards smaller values of $\kappa_*$ at larger
values of $a_{\rm min}$.

The fits for $a_{\rm min}/r_{\rm vdW} \ge 5$ do not provide satisfactory
results, mainly because of significant problems in reproducing the position of
the resonance near 130\,G. The fits for $a_{\rm min}/r_{\rm vdW} \le 4$
(central panel in Fig.~\ref{fig.1stfit}) appear good, but for lower values of
$a_{\rm min}$ the result may be subject to significant finite-range effects. We
therefore consider $a_{\rm min}/r_{\rm vdW} = 4$ to be the best choice. It
gives $\kappa_* = 0.00643(4) \, a_0^{-1}$, based on averaging the results of the
linear and logarithmic fits. The error given here indicates only the
statistical uncertainty, but the dependence of the results on $a_{\rm min}$
suggests additional systematic errors on the order of 10\%.

The dashed horizontal line and the gray-shaded region in
Fig.~\ref{fig.1stBRange} indicate the value of $\kappa_*$ obtained from the
excited-state resonance in Sec.~\ref{sec:excited}, together with the
corresponding error range. It may be seen that our results are consistent with
discrete scaling as described by Eq.~(\ref{eq:kappa_universal}) within the
relatively large uncertainties due to finite-range effects in the low-field
region.

\section{Conclusion}
\label{sec:discuss}

We have reanalyzed experimental results for the Efimov recombination resonances
in $^6$Li arising from the ground and excited Efimov states, using a very
precise model of the two-body scattering \cite{Zurn2013pco} and a new model of
temperature-dependent effects in three-body recombinantion of three
nonidentical fermions. From the excited-state Efimov resonance
\cite{Williams2009efa}, we obtain the value for the 3BP in the wavenumber
representation,
\begin{equation}
\kappa_*  =  0.00678(6)\,a_0^{-1} \,  . \nonumber
\end{equation}
This gives the reduced 3BP
\begin{equation}
\kappa_* r_{\rm vdW}  =  0.212(2)\, . \nonumber
\end{equation}
According to Eq.~(\ref{eq:aminus}) this corresponds to a reduced 3BP in the
scattering length representation,
\begin{equation}
a_-^*/r_{\rm vdW} = -7.11(6) \, . \nonumber
\end{equation}
This latter representation of the 3BP facilitates a direct comparison with
three-boson systems, which are characterized by a single scattering length
\cite{note:aeff}.

Our analysis of the ground-state Efimov resonances \cite{Ottenstein2008cso,
Huckans2009tbr} yields values for the 3BP that are consistent with the above
result within an estimated 10\% uncertainty. Alternatively, they may be viewed
as confirming that the lowest Efimov period in $^6$Li is within 10\% of the
universal value of 22.7. The uncertainties, which follow from systematic shifts
that depend on the choice of the lower cutoff applied to the scattering lengths
in the data analysis, place an upper bound on the magnitude of possible
finite-range corrections to the lowest Efimov period. The rapid decrease of
such shifts with increasing order of the Efimov state \cite{Thogersen2008upo,
Schmidt2012epb} gives us confidence that such corrections can be neglected for
the 3BP if determined from the position of an excited-state resonance.

It is very interesting to compare the present result with the recent
measurement for cesium in Ref.~\cite{Huang2014oot}, which gave
$a_-^{(1)}=-20190(1200)\,a_0$, implying a reduced 3BP $a_-^*/r_{\rm vdW}^{\rm
Cs} = -8.8(4)$ with $r_{\rm vdW}^{\rm Cs} = 101.1\,a_0$
\cite{Berninger2013frw}. In both cases, the Feshbach resonances used for
interaction tuning are strongly entrance-channel-dominated \cite{Chin2010fri}.
The present result for the reduced 3BP in $^6$Li differs from that measured
for Cs by a factor $0.81(4)$. This clearly demonstrates that the van der Waals
length is not the only relevant quantity in determining the 3BP. Even for
strongly entrance-channel-dominated Feshbach resonances, van der Waals
universality of the 3BP is only approximate, and is subject to further
influences. It remains a challenge for theory to understand fully the role of
finite-range effects \cite{Naidon:2011}, of the physics of particular Feshbach
resonances \cite{Schmidt2012epb, Wang2014uvd}, of the role of genuine
short-range three-body forces \cite{Axilrod1943iot, Soldan2003tbn,
Dincao2009tsr}, and of other species-dependent factors such as the number of
bound states in the two-body potentials \cite{Wang2012oot}. It is also possible
that light particles can tunnel through the barrier in the effective potential
\cite{Wang2012oot} more effectively than heavy ones.

It would be highly desirable to investigate other systems at the precision of
the present work, by detecting excited-state Efimov resonances and thus
accurately measuring the 3BP. The bosonic gas of $^7$Li \cite{Gross2009oou,
Pollack2009uit, Rem2013lot} is a prime candidate for such experiments, because
it provides another example of a light system with exceptionally well
characterized two-body scattering properties \cite{Dyke2013frc,
Julienne:Li67:2014}. Atoms such as $^{39}$K \cite{Zaccanti2009ooa, Roy2013tot, Fletcher2013soa}
and $^{85}$Rb \cite{Wild2012mot} also provide very interesting systems for
future precision experiments: $^{85}$Rb offers access to another
entrance-channel-dominated case, while $^7$Li and $^{39}$K offer Feshbach
resonances of intermediate character \cite{Chin2010fri}.


\bigskip
\section*{Acknowledgements}

We thank Selim Jochim for providing the experimental $L_3$ data for the
ground-state Efimov resonances and Benno Rem for useful discussions.
We acknowledge support by the Austrian Science
Fund FWF within project P23106 and by EPSRC under grant no.\ EP/I012044/1. D.S.P. acknowledges support from the IFRAF Institute. The research leading to these results has received funding from the European Research Council (FR7/2007-2013 Grant Agreement no.\ 341197). K.M.O. acknowledges support from the NSF (Grant No. 1312430).


\begin{thebibliography}{68}
\expandafter\ifx\csname natexlab\endcsname\relax\def\natexlab#1{#1}\fi
\expandafter\ifx\csname bibnamefont\endcsname\relax
  \def\bibnamefont#1{#1}\fi
\expandafter\ifx\csname bibfnamefont\endcsname\relax
  \def\bibfnamefont#1{#1}\fi
\expandafter\ifx\csname citenamefont\endcsname\relax
  \def\citenamefont#1{#1}\fi
\expandafter\ifx\csname url\endcsname\relax
  \def\url#1{\texttt{#1}}\fi
\expandafter\ifx\csname urlprefix\endcsname\relax\def\urlprefix{URL }\fi
\providecommand{\bibinfo}[2]{#2}
\providecommand{\eprint}[2][]{\url{#2}}

\bibitem[{\citenamefont{Braaten and Hammer}(2006)}]{Braaten2006uif}
\bibinfo{author}{\bibfnamefont{E.}~\bibnamefont{Braaten}} \bibnamefont{and}
  \bibinfo{author}{\bibfnamefont{H.-W.} \bibnamefont{Hammer}},
  \bibinfo{journal}{Phys. Rep.} \textbf{\bibinfo{volume}{428}},
  \bibinfo{pages}{259} (\bibinfo{year}{2006}).

\bibitem[{\citenamefont{Ferlaino and Grimm}(2010)}]{Ferlaino2010fyo}
\bibinfo{author}{\bibfnamefont{F.}~\bibnamefont{Ferlaino}} \bibnamefont{and}
  \bibinfo{author}{\bibfnamefont{R.}~\bibnamefont{Grimm}},
  \bibinfo{journal}{Physics} \textbf{\bibinfo{volume}{3}}, \bibinfo{pages}{9}
  (\bibinfo{year}{2010}).

\bibitem[{\citenamefont{Efimov}(1970)}]{Efimov1970ela}
\bibinfo{author}{\bibfnamefont{V.}~\bibnamefont{Efimov}},
  \bibinfo{journal}{Phys. Lett. B} \textbf{\bibinfo{volume}{33}},
  \bibinfo{pages}{563} (\bibinfo{year}{1970}).

\bibitem[{not({\natexlab{a}})}]{note:efactor}
\bibinfo{note}{The number $22.7$ is rounded to three digits. The relative
  deviation from the exact value is only $0.025$\%, which is neglible for all
  practical purposes.}

\bibitem[{\citenamefont{D'Incao and Esry}(2006)}]{Dincao2006eto}
\bibinfo{author}{\bibfnamefont{J.~P.} \bibnamefont{D'Incao}} \bibnamefont{and}
  \bibinfo{author}{\bibfnamefont{B.~D.} \bibnamefont{Esry}},
  \bibinfo{journal}{Phys. Rev. A} \textbf{\bibinfo{volume}{73}},
  \bibinfo{pages}{030703(R)} (\bibinfo{year}{2006}).

\bibitem[{\citenamefont{Chin et~al.}(2010)\citenamefont{Chin, Grimm, Julienne,
  and Tiesinga}}]{Chin2010fri}
\bibinfo{author}{\bibfnamefont{C.}~\bibnamefont{Chin}},
  \bibinfo{author}{\bibfnamefont{R.}~\bibnamefont{Grimm}},
  \bibinfo{author}{\bibfnamefont{P.~S.} \bibnamefont{Julienne}},
  \bibnamefont{and} \bibinfo{author}{\bibfnamefont{E.}~\bibnamefont{Tiesinga}},
  \bibinfo{journal}{Rev. Mod. Phys.} \textbf{\bibinfo{volume}{82}},
  \bibinfo{pages}{1225} (\bibinfo{year}{2010}).

\bibitem[{\citenamefont{Kraemer et~al.}(2006)\citenamefont{Kraemer, Mark,
  Waldburger, Danzl, Chin, Engeser, Lange, Pilch, Jaakkola, N\"agerl
  et~al.}}]{Kraemer2006efe}
\bibinfo{author}{\bibfnamefont{T.}~\bibnamefont{Kraemer}},
  \bibinfo{author}{\bibfnamefont{M.}~\bibnamefont{Mark}},
  \bibinfo{author}{\bibfnamefont{P.}~\bibnamefont{Waldburger}},
  \bibinfo{author}{\bibfnamefont{J.~G.} \bibnamefont{Danzl}},
  \bibinfo{author}{\bibfnamefont{C.}~\bibnamefont{Chin}},
  \bibinfo{author}{\bibfnamefont{B.}~\bibnamefont{Engeser}},
  \bibinfo{author}{\bibfnamefont{A.~D.} \bibnamefont{Lange}},
  \bibinfo{author}{\bibfnamefont{K.}~\bibnamefont{Pilch}},
  \bibinfo{author}{\bibfnamefont{A.}~\bibnamefont{Jaakkola}},
  \bibinfo{author}{\bibfnamefont{H.-C.} \bibnamefont{N\"agerl}}, \bibnamefont{and}
  \bibinfo{author}{\bibfnamefont{R.} \bibnamefont{Grimm}},
  \bibinfo{journal}{Nature (London)}
  \textbf{\bibinfo{volume}{440}}, \bibinfo{pages}{315} (\bibinfo{year}{2006}).

\bibitem[{\citenamefont{Zaccanti et~al.}(2009)\citenamefont{Zaccanti, Deissler,
  D'Errico, Fattori, Jona-Lasinio, M\"uller, Roati, Inguscio, and
  Modugno}}]{Zaccanti2009ooa}
\bibinfo{author}{\bibfnamefont{M.}~\bibnamefont{Zaccanti}},
  \bibinfo{author}{\bibfnamefont{B.}~\bibnamefont{Deissler}},
  \bibinfo{author}{\bibfnamefont{C.}~\bibnamefont{D'Errico}},
  \bibinfo{author}{\bibfnamefont{M.}~\bibnamefont{Fattori}},
  \bibinfo{author}{\bibfnamefont{M.}~\bibnamefont{Jona-Lasinio}},
  \bibinfo{author}{\bibfnamefont{S.}~\bibnamefont{M\"uller}},
  \bibinfo{author}{\bibfnamefont{G.}~\bibnamefont{Roati}},
  \bibinfo{author}{\bibfnamefont{M.}~\bibnamefont{Inguscio}}, \bibnamefont{and}
  \bibinfo{author}{\bibfnamefont{G.}~\bibnamefont{Modugno}},
  \bibinfo{journal}{Nat. Phys.} \textbf{\bibinfo{volume}{5}},
  \bibinfo{pages}{586} (\bibinfo{year}{2009}).

\bibitem[{\citenamefont{Pollack et~al.}(2009)\citenamefont{Pollack, Dries, and
  Hulet}}]{Pollack2009uit}
\bibinfo{author}{\bibfnamefont{S.~E.} \bibnamefont{Pollack}},
  \bibinfo{author}{\bibfnamefont{D.}~\bibnamefont{Dries}}, \bibnamefont{and}
  \bibinfo{author}{\bibfnamefont{R.~G.} \bibnamefont{Hulet}},
  \bibinfo{journal}{Science} \textbf{\bibinfo{volume}{326}},
  \bibinfo{pages}{1683} (\bibinfo{year}{2009}).

\bibitem[{\citenamefont{Gross et~al.}(2009)\citenamefont{Gross, Shotan,
  Kokkelmans, and Khaykovich}}]{Gross2009oou}
\bibinfo{author}{\bibfnamefont{N.}~\bibnamefont{Gross}},
  \bibinfo{author}{\bibfnamefont{Z.}~\bibnamefont{Shotan}},
  \bibinfo{author}{\bibfnamefont{S.}~\bibnamefont{Kokkelmans}},
  \bibnamefont{and}
  \bibinfo{author}{\bibfnamefont{L.}~\bibnamefont{Khaykovich}},
  \bibinfo{journal}{Phys. Rev. Lett.} \textbf{\bibinfo{volume}{103}},
  \bibinfo{pages}{163202} (\bibinfo{year}{2009}).

\bibitem[{\citenamefont{Gross et~al.}(2010)\citenamefont{Gross, Shotan,
  Kokkelmans, and Khaykovich}}]{Gross2010nsi}
\bibinfo{author}{\bibfnamefont{N.}~\bibnamefont{Gross}},
  \bibinfo{author}{\bibfnamefont{Z.}~\bibnamefont{Shotan}},
  \bibinfo{author}{\bibfnamefont{S.}~\bibnamefont{Kokkelmans}},
  \bibnamefont{and}
  \bibinfo{author}{\bibfnamefont{L.}~\bibnamefont{Khaykovich}},
  \bibinfo{journal}{Phys. Rev. Lett} \textbf{\bibinfo{volume}{105}},
  \bibinfo{pages}{103203} (\bibinfo{year}{2010}).

\bibitem[{\citenamefont{Wild et~al.}(2012)\citenamefont{Wild, Makotyn, Pino,
  Cornell, and Jin}}]{Wild2012mot}
\bibinfo{author}{\bibfnamefont{R.~J.} \bibnamefont{Wild}},
  \bibinfo{author}{\bibfnamefont{P.}~\bibnamefont{Makotyn}},
  \bibinfo{author}{\bibfnamefont{J.~M.} \bibnamefont{Pino}},
  \bibinfo{author}{\bibfnamefont{E.~A.} \bibnamefont{Cornell}},
  \bibnamefont{and} \bibinfo{author}{\bibfnamefont{D.~S.} \bibnamefont{Jin}},
  \bibinfo{journal}{Phys. Rev. Lett.} \textbf{\bibinfo{volume}{108}},
  \bibinfo{pages}{145305} (\bibinfo{year}{2012}).

\bibitem[{\citenamefont{Roy et~al.}(2013)\citenamefont{Roy, Landini,
  Trenkwalder, Semeghini, Spagnolli, Simoni, Fattori, Inguscio, and
  Modugno}}]{Roy2013tot}
\bibinfo{author}{\bibfnamefont{S.}~\bibnamefont{Roy}},
  \bibinfo{author}{\bibfnamefont{M.}~\bibnamefont{Landini}},
  \bibinfo{author}{\bibfnamefont{A.}~\bibnamefont{Trenkwalder}},
  \bibinfo{author}{\bibfnamefont{G.}~\bibnamefont{Semeghini}},
  \bibinfo{author}{\bibfnamefont{G.}~\bibnamefont{Spagnolli}},
  \bibinfo{author}{\bibfnamefont{A.}~\bibnamefont{Simoni}},
  \bibinfo{author}{\bibfnamefont{M.}~\bibnamefont{Fattori}},
  \bibinfo{author}{\bibfnamefont{M.}~\bibnamefont{Inguscio}}, \bibnamefont{and}
  \bibinfo{author}{\bibfnamefont{G.}~\bibnamefont{Modugno}},
  \bibinfo{journal}{Phys. Rev. Lett.} \textbf{\bibinfo{volume}{111}},
  \bibinfo{pages}{053202} (\bibinfo{year}{2013}).

\bibitem[{\citenamefont{Ottenstein et~al.}(2008)\citenamefont{Ottenstein,
  Lompe, Kohnen, Wenz, and Jochim}}]{Ottenstein2008cso}
\bibinfo{author}{\bibfnamefont{T.~B.} \bibnamefont{Ottenstein}},
  \bibinfo{author}{\bibfnamefont{T.}~\bibnamefont{Lompe}},
  \bibinfo{author}{\bibfnamefont{M.}~\bibnamefont{Kohnen}},
  \bibinfo{author}{\bibfnamefont{A.~N.} \bibnamefont{Wenz}}, \bibnamefont{and}
  \bibinfo{author}{\bibfnamefont{S.}~\bibnamefont{Jochim}},
  \bibinfo{journal}{Phys. Rev. Lett.} \textbf{\bibinfo{volume}{101}},
  \bibinfo{eid}{203202} (\bibinfo{year}{2008}).

\bibitem[{\citenamefont{Huckans et~al.}(2009)\citenamefont{Huckans, Williams,
  Hazlett, Stites, and O'Hara}}]{Huckans2009tbr}
\bibinfo{author}{\bibfnamefont{J.~H.} \bibnamefont{Huckans}},
  \bibinfo{author}{\bibfnamefont{J.~R.} \bibnamefont{Williams}},
  \bibinfo{author}{\bibfnamefont{E.~L.} \bibnamefont{Hazlett}},
  \bibinfo{author}{\bibfnamefont{R.~W.} \bibnamefont{Stites}},
  \bibnamefont{and} \bibinfo{author}{\bibfnamefont{K.~M.}
  \bibnamefont{O'Hara}}, \bibinfo{journal}{Phys. Rev. Lett.}
  \textbf{\bibinfo{volume}{102}}, \bibinfo{eid}{165302} (\bibinfo{year}{2009}).

\bibitem[{\citenamefont{Williams et~al.}(2009)\citenamefont{Williams, Hazlett,
  Huckans, Stites, Zhang, and O'Hara}}]{Williams2009efa}
\bibinfo{author}{\bibfnamefont{J.~R.} \bibnamefont{Williams}},
  \bibinfo{author}{\bibfnamefont{E.~L.} \bibnamefont{Hazlett}},
  \bibinfo{author}{\bibfnamefont{J.~H.} \bibnamefont{Huckans}},
  \bibinfo{author}{\bibfnamefont{R.~W.} \bibnamefont{Stites}},
  \bibinfo{author}{\bibfnamefont{Y.}~\bibnamefont{Zhang}}, \bibnamefont{and}
  \bibinfo{author}{\bibfnamefont{K.~M.} \bibnamefont{O'Hara}},
  \bibinfo{journal}{Phys. Rev. Lett.} \textbf{\bibinfo{volume}{103}},
  \bibinfo{pages}{130404} (\bibinfo{year}{2009}).

\bibitem[{\citenamefont{Nakajima et~al.}(2010)\citenamefont{Nakajima,
  Horikoshi, Mukaiyama, Naidon, and Ueda}}]{Nakajima2010nea}
\bibinfo{author}{\bibfnamefont{S.}~\bibnamefont{Nakajima}},
  \bibinfo{author}{\bibfnamefont{M.}~\bibnamefont{Horikoshi}},
  \bibinfo{author}{\bibfnamefont{T.}~\bibnamefont{Mukaiyama}},
  \bibinfo{author}{\bibfnamefont{P.}~\bibnamefont{Naidon}}, \bibnamefont{and}
  \bibinfo{author}{\bibfnamefont{M.}~\bibnamefont{Ueda}},
  \bibinfo{journal}{Phys. Rev. Lett} \textbf{\bibinfo{volume}{105}},
  \bibinfo{pages}{023201} (\bibinfo{year}{2010}).

\bibitem[{\citenamefont{Barontini et~al.}(2009)\citenamefont{Barontini, Weber,
  Rabatti, Catani, Thalhammer, Inguscio, and Minardi}}]{Barontini2009ooh}
\bibinfo{author}{\bibfnamefont{G.}~\bibnamefont{Barontini}},
  \bibinfo{author}{\bibfnamefont{C.}~\bibnamefont{Weber}},
  \bibinfo{author}{\bibfnamefont{F.}~\bibnamefont{Rabatti}},
  \bibinfo{author}{\bibfnamefont{J.}~\bibnamefont{Catani}},
  \bibinfo{author}{\bibfnamefont{G.}~\bibnamefont{Thalhammer}},
  \bibinfo{author}{\bibfnamefont{M.}~\bibnamefont{Inguscio}}, \bibnamefont{and}
  \bibinfo{author}{\bibfnamefont{F.}~\bibnamefont{Minardi}},
  \bibinfo{journal}{Phys. Rev. Lett.} \textbf{\bibinfo{volume}{103}},
  \bibinfo{pages}{043201} (\bibinfo{year}{2009}).

\bibitem[{\citenamefont{Bloom et~al.}(2013)\citenamefont{Bloom, Hu, Cumby, and
  Jin}}]{Bloom2013tou}
\bibinfo{author}{\bibfnamefont{R.~S.} \bibnamefont{Bloom}},
  \bibinfo{author}{\bibfnamefont{M.-G.} \bibnamefont{Hu}},
  \bibinfo{author}{\bibfnamefont{T.~D.} \bibnamefont{Cumby}}, \bibnamefont{and}
  \bibinfo{author}{\bibfnamefont{D.~S.} \bibnamefont{Jin}},
  \bibinfo{journal}{Phys. Rev. Lett.} \textbf{\bibinfo{volume}{111}},
  \bibinfo{pages}{105301} (\bibinfo{year}{2013}).

\bibitem[{\citenamefont{Tung et~al.}(2014)\citenamefont{Tung, Jimenez-Garcia,
  Johansen, Parker, and Chin}}]{Tung2014oog}
\bibinfo{author}{\bibfnamefont{S.-K.} \bibnamefont{Tung}},
  \bibinfo{author}{\bibfnamefont{K.}~\bibnamefont{Jimenez-Garcia}},
  \bibinfo{author}{\bibfnamefont{J.}~\bibnamefont{Johansen}},
  \bibinfo{author}{\bibfnamefont{C.~V.} \bibnamefont{Parker}},
  \bibnamefont{and} \bibinfo{author}{\bibfnamefont{C.}~\bibnamefont{Chin}},
  \bibinfo{journal}{arXiv:1402.5943}  (\bibinfo{year}{2014}).

\bibitem[{\citenamefont{Pires et~al.}(2014)\citenamefont{Pires, Ulmanis,
  H\"afner, Repp, Arias, Kuhnle, and Weidem\"uller}}]{Pires2014ooe}
\bibinfo{author}{\bibfnamefont{R.}~\bibnamefont{Pires}},
  \bibinfo{author}{\bibfnamefont{J.}~\bibnamefont{Ulmanis}},
  \bibinfo{author}{\bibfnamefont{S.}~\bibnamefont{H\"afner}},
  \bibinfo{author}{\bibfnamefont{M.}~\bibnamefont{Repp}},
  \bibinfo{author}{\bibfnamefont{A.}~\bibnamefont{Arias}},
  \bibinfo{author}{\bibfnamefont{E.~D.} \bibnamefont{Kuhnle}},
  \bibnamefont{and}
  \bibinfo{author}{\bibfnamefont{M.}~\bibnamefont{Weidem\"uller}},
  \bibinfo{journal}{Phys. Rev. Lett.} \textbf{\bibinfo{volume}{112}},
  \bibinfo{pages}{250404} (\bibinfo{year}{2014}).

\bibitem[{\citenamefont{Hammer and Platter}(2007)}]{Hammer2007upo}
\bibinfo{author}{\bibfnamefont{H.-W.} \bibnamefont{Hammer}} \bibnamefont{and}
  \bibinfo{author}{\bibfnamefont{L.}~\bibnamefont{Platter}},
  \bibinfo{journal}{Eur. Phys. J. A} \textbf{\bibinfo{volume}{32}},
  \bibinfo{pages}{113} (\bibinfo{year}{2007}).

\bibitem[{\citenamefont{{von Stecher} et~al.}(2009)\citenamefont{{von Stecher},
  D'Incao, and Greene}}]{vonStecher2009sou}
\bibinfo{author}{\bibfnamefont{J.}~\bibnamefont{{von Stecher}}},
  \bibinfo{author}{\bibfnamefont{J.~P.} \bibnamefont{D'Incao}},
  \bibnamefont{and} \bibinfo{author}{\bibfnamefont{C.~H.}
  \bibnamefont{Greene}}, \bibinfo{journal}{Nat. Phys.}
  \textbf{\bibinfo{volume}{5}}, \bibinfo{pages}{417} (\bibinfo{year}{2009}).

\bibitem[{\citenamefont{{von Stecher}}(2010)}]{vonStecher2010wbc}
\bibinfo{author}{\bibfnamefont{J.}~\bibnamefont{{von Stecher}}},
  \bibinfo{journal}{J. Phys. B} \textbf{\bibinfo{volume}{43}},
  \bibinfo{pages}{101002} (\bibinfo{year}{2010}).

\bibitem[{\citenamefont{Ferlaino et~al.}(2009)\citenamefont{Ferlaino, Knoop,
  Berninger, Harm, {D'Incao}, N\"agerl, and Grimm}}]{Ferlaino2009efu}
\bibinfo{author}{\bibfnamefont{F.}~\bibnamefont{Ferlaino}},
  \bibinfo{author}{\bibfnamefont{S.}~\bibnamefont{Knoop}},
  \bibinfo{author}{\bibfnamefont{M.}~\bibnamefont{Berninger}},
  \bibinfo{author}{\bibfnamefont{W.}~\bibnamefont{Harm}},
  \bibinfo{author}{\bibfnamefont{J.~P.} \bibnamefont{{D'Incao}}},
  \bibinfo{author}{\bibfnamefont{H.-C.} \bibnamefont{N\"agerl}},
  \bibnamefont{and} \bibinfo{author}{\bibfnamefont{R.}~\bibnamefont{Grimm}},
  \bibinfo{journal}{Phys. Rev. Lett.} \textbf{\bibinfo{volume}{102}},
  \bibinfo{pages}{140401} (\bibinfo{year}{2009}).

\bibitem[{\citenamefont{Zenesini et~al.}(2013)\citenamefont{Zenesini, Huang,
  Berninger, Besler, N\"{a}gerl, Ferlaino, Grimm, Greene, and von
  Stecher}}]{Zenesini2013rfb}
\bibinfo{author}{\bibfnamefont{A.}~\bibnamefont{Zenesini}},
  \bibinfo{author}{\bibfnamefont{B.}~\bibnamefont{Huang}},
  \bibinfo{author}{\bibfnamefont{M.}~\bibnamefont{Berninger}},
  \bibinfo{author}{\bibfnamefont{S.}~\bibnamefont{Besler}},
  \bibinfo{author}{\bibfnamefont{H.-C.} \bibnamefont{N\"{a}gerl}},
  \bibinfo{author}{\bibfnamefont{F.}~\bibnamefont{Ferlaino}},
  \bibinfo{author}{\bibfnamefont{R.}~\bibnamefont{Grimm}},
  \bibinfo{author}{\bibfnamefont{C.~H.} \bibnamefont{Greene}},
  \bibnamefont{and} \bibinfo{author}{\bibfnamefont{J.}~\bibnamefont{von
  Stecher}}, \bibinfo{journal}{New J. Phys.} \textbf{\bibinfo{volume}{15}},
  \bibinfo{pages}{043040} (\bibinfo{year}{2013}).

\bibitem[{\citenamefont{Berninger et~al.}(2011)\citenamefont{Berninger,
  Zenesini, Huang, Harm, N\"{a}gerl, Ferlaino, Grimm, Julienne, and
  Hutson}}]{Berninger2011uot}
\bibinfo{author}{\bibfnamefont{M.}~\bibnamefont{Berninger}},
  \bibinfo{author}{\bibfnamefont{A.}~\bibnamefont{Zenesini}},
  \bibinfo{author}{\bibfnamefont{B.}~\bibnamefont{Huang}},
  \bibinfo{author}{\bibfnamefont{W.}~\bibnamefont{Harm}},
  \bibinfo{author}{\bibfnamefont{H.-C.} \bibnamefont{N\"{a}gerl}},
  \bibinfo{author}{\bibfnamefont{F.}~\bibnamefont{Ferlaino}},
  \bibinfo{author}{\bibfnamefont{R.}~\bibnamefont{Grimm}},
  \bibinfo{author}{\bibfnamefont{P.~S.} \bibnamefont{Julienne}},
  \bibnamefont{and} \bibinfo{author}{\bibfnamefont{J.~M.}
  \bibnamefont{Hutson}}, \bibinfo{journal}{Phys. Rev. Lett.}
  \textbf{\bibinfo{volume}{107}}, \bibinfo{pages}{120401}
  (\bibinfo{year}{2011}).

\bibitem[{\citenamefont{Chin}(2011)}]{Chin2011uso}
\bibinfo{author}{\bibfnamefont{C.}~\bibnamefont{Chin}},
  \bibinfo{journal}{arXiv:1111.1484}  (\bibinfo{year}{2011}).

\bibitem[{\citenamefont{Wang et~al.}(2012)\citenamefont{Wang, D'Incao, Esry,
  and Greene}}]{Wang2012oot}
\bibinfo{author}{\bibfnamefont{J.}~\bibnamefont{Wang}},
  \bibinfo{author}{\bibfnamefont{J.~P.} \bibnamefont{D'Incao}},
  \bibinfo{author}{\bibfnamefont{B.~D.} \bibnamefont{Esry}}, \bibnamefont{and}
  \bibinfo{author}{\bibfnamefont{C.~H.} \bibnamefont{Greene}},
  \bibinfo{journal}{Phys. Rev. Lett.} \textbf{\bibinfo{volume}{108}},
  \bibinfo{pages}{263001} (\bibinfo{year}{2012}).

\bibitem[{\citenamefont{Schmidt et~al.}(2012)\citenamefont{Schmidt, Rath, and
  Zwerger}}]{Schmidt2012epb}
\bibinfo{author}{\bibfnamefont{R.}~\bibnamefont{Schmidt}},
  \bibinfo{author}{\bibfnamefont{S.}~\bibnamefont{Rath}}, \bibnamefont{and}
  \bibinfo{author}{\bibfnamefont{W.}~\bibnamefont{Zwerger}},
  \bibinfo{journal}{Eur. Phys. J. B} \textbf{\bibinfo{volume}{85}},
  \bibinfo{eid}{386} (\bibinfo{year}{2012}).
  
\bibitem[{\citenamefont{S\o{}rensen et~al.}(2012)\citenamefont{S\o{}rensen,
  Fedorov, Jensen, and Zinner}}]{Sorensen2012epa}
\bibinfo{author}{\bibfnamefont{P.~K.} \bibnamefont{S\o{}rensen}},
  \bibinfo{author}{\bibfnamefont{D.~V.} \bibnamefont{Fedorov}},
  \bibinfo{author}{\bibfnamefont{A.~S.} \bibnamefont{Jensen}},
  \bibnamefont{and} \bibinfo{author}{\bibfnamefont{N.~T.}
  \bibnamefont{Zinner}}, \bibinfo{journal}{Phys. Rev. A}
  \textbf{\bibinfo{volume}{86}}, \bibinfo{pages}{052516}
  (\bibinfo{year}{2012}).

\bibitem[{\citenamefont{Naidon et~al.}(2014)\citenamefont{Naidon, Endo, and
  Ueda}}]{Naidon2014moa}
\bibinfo{author}{\bibfnamefont{P.}~\bibnamefont{Naidon}},
  \bibinfo{author}{\bibfnamefont{S.}~\bibnamefont{Endo}}, \bibnamefont{and}
  \bibinfo{author}{\bibfnamefont{M.}~\bibnamefont{Ueda}},
  \bibinfo{journal}{Phys. Rev. Lett.} \textbf{\bibinfo{volume}{112}},
  \bibinfo{pages}{105301} (\bibinfo{year}{2014}).

\bibitem[{\citenamefont{Wang and Julienne}(2014)}]{Wang2014uvd}
\bibinfo{author}{\bibfnamefont{Y.}~\bibnamefont{Wang}} \bibnamefont{and}
  \bibinfo{author}{\bibfnamefont{P.~S.} \bibnamefont{Julienne}},
  \bibinfo{journal}{Nat. Phys.} \textbf{\bibinfo{volume}{10}},
  \bibinfo{pages}{768} (\bibinfo{year}{2014}).

\bibitem[{\citenamefont{Esry et~al.}(1999)\citenamefont{Esry, Greene, and
  Burke}}]{Esry1999rot}
\bibinfo{author}{\bibfnamefont{B.~D.} \bibnamefont{Esry}},
  \bibinfo{author}{\bibfnamefont{C.~H.} \bibnamefont{Greene}},
  \bibnamefont{and} \bibinfo{author}{\bibfnamefont{J.~P.} \bibnamefont{Burke}},
  \bibinfo{journal}{Phys. Rev. Lett.} \textbf{\bibinfo{volume}{83}},
  \bibinfo{pages}{1751} (\bibinfo{year}{1999}).

\bibitem[{\citenamefont{Braaten and Hammer}(2001)}]{Braaten2001tbr}
\bibinfo{author}{\bibfnamefont{E.}~\bibnamefont{Braaten}} \bibnamefont{and}
  \bibinfo{author}{\bibfnamefont{H.-W.} \bibnamefont{Hammer}},
  \bibinfo{journal}{Phys. Rev. Lett.} \textbf{\bibinfo{volume}{87}},
  \bibinfo{pages}{160407} (\bibinfo{year}{2001}).

\bibitem[{\citenamefont{Ferlaino et~al.}(2011)\citenamefont{Ferlaino, Zenesini,
  Berninger, Huang, N\"{a}gerl, and Grimm}}]{Ferlaino2011eri}
\bibinfo{author}{\bibfnamefont{F.}~\bibnamefont{Ferlaino}},
  \bibinfo{author}{\bibfnamefont{A.}~\bibnamefont{Zenesini}},
  \bibinfo{author}{\bibfnamefont{M.}~\bibnamefont{Berninger}},
  \bibinfo{author}{\bibfnamefont{B.}~\bibnamefont{Huang}},
  \bibinfo{author}{\bibfnamefont{H.-C.} \bibnamefont{N\"{a}gerl}},
  \bibnamefont{and} \bibinfo{author}{\bibfnamefont{R.}~\bibnamefont{Grimm}},
  \bibinfo{journal}{Few-Body Syst.} \textbf{\bibinfo{volume}{51}},
  \bibinfo{pages}{113} (\bibinfo{year}{2011}).

\bibitem[{\citenamefont{Th{\o}gersen
  et~al.}(2008{\natexlab{a}})\citenamefont{Th{\o}gersen, Fedorov, and
  Jensen}}]{Thogersen2008nbe}
\bibinfo{author}{\bibfnamefont{M.}~\bibnamefont{Th{\o}gersen}},
  \bibinfo{author}{\bibfnamefont{D.~V.} \bibnamefont{Fedorov}},
  \bibnamefont{and} \bibinfo{author}{\bibfnamefont{A.~S.}
  \bibnamefont{Jensen}}, \bibinfo{journal}{Europhys. Lett.}
  \textbf{\bibinfo{volume}{83}}, \bibinfo{pages}{30012}
  (\bibinfo{year}{2008}{\natexlab{a}}).

\bibitem[{\citenamefont{Platter et~al.}(2009)\citenamefont{Platter, Ji, and
  Phillips}}]{Platter2009rct}
\bibinfo{author}{\bibfnamefont{L.}~\bibnamefont{Platter}},
  \bibinfo{author}{\bibfnamefont{C.}~\bibnamefont{Ji}}, \bibnamefont{and}
  \bibinfo{author}{\bibfnamefont{D.~R.} \bibnamefont{Phillips}},
  \bibinfo{journal}{Phys. Rev. A} \textbf{\bibinfo{volume}{79}},
  \bibinfo{pages}{022702} (\bibinfo{year}{2009}).

\bibitem[{\citenamefont{Naidon and Ueda}(2011)}]{Naidon:2011}
\bibinfo{author}{\bibfnamefont{P.}~\bibnamefont{Naidon}} \bibnamefont{and}
  \bibinfo{author}{\bibfnamefont{M.}~\bibnamefont{Ueda}},
  \bibinfo{journal}{C. R. Phys.} \textbf{\bibinfo{volume}{12}},
  \bibinfo{pages}{13} (\bibinfo{year}{2011}).

\bibitem[{\citenamefont{D'Incao et~al.}(2004)\citenamefont{D'Incao, Suno, and
  Esry}}]{Dincao2004lou}
\bibinfo{author}{\bibfnamefont{J.~P.} \bibnamefont{D'Incao}},
  \bibinfo{author}{\bibfnamefont{H.}~\bibnamefont{Suno}}, \bibnamefont{and}
  \bibinfo{author}{\bibfnamefont{B.~D.} \bibnamefont{Esry}},
  \bibinfo{journal}{Phys. Rev. Lett.} \textbf{\bibinfo{volume}{93}},
  \bibinfo{pages}{123201} (\bibinfo{year}{2004}).

\bibitem[{\citenamefont{Rem et~al.}(2013)\citenamefont{Rem, Grier,
  Ferrier-Barbut, Eismann, Langen, Navon, Khaykovich, Werner, Petrov, Chevy
  et~al.}}]{Rem2013lot}
\bibinfo{author}{\bibfnamefont{B.~S.} \bibnamefont{Rem}},
  \bibinfo{author}{\bibfnamefont{A.~T.} \bibnamefont{Grier}},
  \bibinfo{author}{\bibfnamefont{I.}~\bibnamefont{Ferrier-Barbut}},
  \bibinfo{author}{\bibfnamefont{U.}~\bibnamefont{Eismann}},
  \bibinfo{author}{\bibfnamefont{T.}~\bibnamefont{Langen}},
  \bibinfo{author}{\bibfnamefont{N.}~\bibnamefont{Navon}},
  \bibinfo{author}{\bibfnamefont{L.}~\bibnamefont{Khaykovich}},
  \bibinfo{author}{\bibfnamefont{F.}~\bibnamefont{Werner}},
  \bibinfo{author}{\bibfnamefont{D.~S.} \bibnamefont{Petrov}},
  \bibinfo{author}{\bibfnamefont{F.}~\bibnamefont{Chevy}}, \bibnamefont{and}
  \bibinfo{author}{\bibfnamefont{C.} \bibnamefont{Salomon}}, 
  \bibinfo{journal}{Phys. Rev. Lett.}
  \textbf{\bibinfo{volume}{110}}, \bibinfo{pages}{163202}
  (\bibinfo{year}{2013}).

\bibitem[{\citenamefont{Huang et~al.}(2014)\citenamefont{Huang, Sidorenkov,
  Grimm, and Hutson}}]{Huang2014oot}
\bibinfo{author}{\bibfnamefont{B.}~\bibnamefont{Huang}},
  \bibinfo{author}{\bibfnamefont{L.~A.}~\bibnamefont{Sidorenkov}},
  \bibinfo{author}{\bibfnamefont{R.}~\bibnamefont{Grimm}}, \bibnamefont{and}
  \bibinfo{author}{\bibfnamefont{J.~M.} \bibnamefont{Hutson}},
  \bibinfo{journal}{Phys. Rev. Lett.} \textbf{\bibinfo{volume}{112}},
  \bibinfo{pages}{190401} (\bibinfo{year}{2014}).

\bibitem[{\citenamefont{Berninger et~al.}(2013)\citenamefont{Berninger,
  Zenesini, Huang, Harm, N\"agerl, Ferlaino, Grimm, Julienne, and
  Hutson}}]{Berninger2013frw}
\bibinfo{author}{\bibfnamefont{M.}~\bibnamefont{Berninger}},
  \bibinfo{author}{\bibfnamefont{A.}~\bibnamefont{Zenesini}},
  \bibinfo{author}{\bibfnamefont{B.}~\bibnamefont{Huang}},
  \bibinfo{author}{\bibfnamefont{W.}~\bibnamefont{Harm}},
  \bibinfo{author}{\bibfnamefont{H.-C.} \bibnamefont{N\"agerl}},
  \bibinfo{author}{\bibfnamefont{F.}~\bibnamefont{Ferlaino}},
  \bibinfo{author}{\bibfnamefont{R.}~\bibnamefont{Grimm}},
  \bibinfo{author}{\bibfnamefont{P.~S.} \bibnamefont{Julienne}},
  \bibnamefont{and} \bibinfo{author}{\bibfnamefont{J.~M.}
  \bibnamefont{Hutson}}, \bibinfo{journal}{Phys. Rev. A}
  \textbf{\bibinfo{volume}{87}}, \bibinfo{pages}{032517}
  (\bibinfo{year}{2013}).

\bibitem[{\citenamefont{Zhen and Macek}(1986)}]{Macek:1986}
\bibinfo{author}{\bibfnamefont{Z.}~\bibnamefont{Zhen}} \bibnamefont{and}
  \bibinfo{author}{\bibfnamefont{J.}~\bibnamefont{Macek}}, \bibinfo{journal}{Z.
  Phys. D} \textbf{\bibinfo{volume}{3}}, \bibinfo{pages}{31}
  (\bibinfo{year}{1986}).

\bibitem[{\citenamefont{Wenz et~al.}(2009)\citenamefont{Wenz, Lompe,
  Ottenstein, Serwane, Z\"urn, and Jochim}}]{Wenz2009uti}
\bibinfo{author}{\bibfnamefont{A.~N.} \bibnamefont{Wenz}},
  \bibinfo{author}{\bibfnamefont{T.}~\bibnamefont{Lompe}},
  \bibinfo{author}{\bibfnamefont{T.~B.} \bibnamefont{Ottenstein}},
  \bibinfo{author}{\bibfnamefont{F.}~\bibnamefont{Serwane}},
  \bibinfo{author}{\bibfnamefont{G.}~\bibnamefont{Z\"urn}}, \bibnamefont{and}
  \bibinfo{author}{\bibfnamefont{S.}~\bibnamefont{Jochim}},
  \bibinfo{journal}{Phys. Rev. A} \textbf{\bibinfo{volume}{80}},
  \bibinfo{pages}{040702(R)} (\bibinfo{year}{2009}).

\bibitem[{\citenamefont{Braaten et~al.}(2009)\citenamefont{Braaten, Hammer,
  Kang, and Platter}}]{Braaten2009tbr}
\bibinfo{author}{\bibfnamefont{E.}~\bibnamefont{Braaten}},
  \bibinfo{author}{\bibfnamefont{H.-W.} \bibnamefont{Hammer}},
  \bibinfo{author}{\bibfnamefont{D.}~\bibnamefont{Kang}}, \bibnamefont{and}
  \bibinfo{author}{\bibfnamefont{L.}~\bibnamefont{Platter}},
  \bibinfo{journal}{Phys. Rev. Lett.} \textbf{\bibinfo{volume}{103}},
  \bibinfo{pages}{073202} (\bibinfo{year}{2009}).

\bibitem[{\citenamefont{Esry et~al.}(2001)\citenamefont{Esry, Greene, and
  Suno}}]{Esry2001tlf}
\bibinfo{author}{\bibfnamefont{B.~D.} \bibnamefont{Esry}},
  \bibinfo{author}{\bibfnamefont{C.~H.} \bibnamefont{Greene}},
  \bibnamefont{and} \bibinfo{author}{\bibfnamefont{H.}~\bibnamefont{Suno}},
  \bibinfo{journal}{Phys. Rev. A} \textbf{\bibinfo{volume}{65}},
  \bibinfo{pages}{010705} (\bibinfo{year}{2001}).

\bibitem[{\citenamefont{Z\"urn et~al.}(2013)\citenamefont{Z\"urn, Lompe, Wenz,
  Jochim, Julienne, and Hutson}}]{Zurn2013pco}
\bibinfo{author}{\bibfnamefont{G.}~\bibnamefont{Z\"urn}},
  \bibinfo{author}{\bibfnamefont{T.}~\bibnamefont{Lompe}},
  \bibinfo{author}{\bibfnamefont{A.~N.} \bibnamefont{Wenz}},
  \bibinfo{author}{\bibfnamefont{S.}~\bibnamefont{Jochim}},
  \bibinfo{author}{\bibfnamefont{P.~S.} \bibnamefont{Julienne}},
  \bibnamefont{and} \bibinfo{author}{\bibfnamefont{J.~M.}
  \bibnamefont{Hutson}}, \bibinfo{journal}{Phys. Rev. Lett.}
  \textbf{\bibinfo{volume}{110}}, \bibinfo{pages}{135301}
  (\bibinfo{year}{2013}).

\bibitem[{\citenamefont{Bartenstein et~al.}(2005)\citenamefont{Bartenstein,
  Altmeyer, Riedl, Geursen, Jochim, Chin, {Hecker Denschlag}, Grimm, Simoni,
  Tiesinga et~al.}}]{Bartenstein2005pdo}
\bibinfo{author}{\bibfnamefont{M.}~\bibnamefont{Bartenstein}},
  \bibinfo{author}{\bibfnamefont{A.}~\bibnamefont{Altmeyer}},
  \bibinfo{author}{\bibfnamefont{S.}~\bibnamefont{Riedl}},
  \bibinfo{author}{\bibfnamefont{R.}~\bibnamefont{Geursen}},
  \bibinfo{author}{\bibfnamefont{S.}~\bibnamefont{Jochim}},
  \bibinfo{author}{\bibfnamefont{C.}~\bibnamefont{Chin}},
  \bibinfo{author}{\bibfnamefont{J.}~\bibnamefont{{Hecker Denschlag}}},
  \bibinfo{author}{\bibfnamefont{R.}~\bibnamefont{Grimm}},
  \bibinfo{author}{\bibfnamefont{A.}~\bibnamefont{Simoni}},
  \bibinfo{author}{\bibfnamefont{E.}~\bibnamefont{Tiesinga}}, 
  \bibinfo{author}{\bibfnamefont{C.~J.} \bibnamefont{Williams}}, \bibnamefont{and} 
  \bibinfo{author}{\bibfnamefont{P.~S.} \bibnamefont{Julienne}},
  \bibinfo{journal}{Phys. Rev. Lett.}
  \textbf{\bibinfo{volume}{94}}, \bibinfo{pages}{103201}
  (\bibinfo{year}{2005}).

\bibitem[{\citenamefont{Bethe}(1949)}]{Bethe:1949}
\bibinfo{author}{\bibfnamefont{H.~A.} \bibnamefont{Bethe}},
  \bibinfo{journal}{Phys. Rev.} \textbf{\bibinfo{volume}{76}},
  \bibinfo{pages}{38} (\bibinfo{year}{1949}).

\bibitem[{\citenamefont{Hinckelmann and Spruch}(1971)}]{Hinckelmann:1971}
\bibinfo{author}{\bibfnamefont{O.}~\bibnamefont{Hinckelmann}} \bibnamefont{and}
  \bibinfo{author}{\bibfnamefont{L.}~\bibnamefont{Spruch}},
  \bibinfo{journal}{Phys. Rev. A} \textbf{\bibinfo{volume}{3}},
  \bibinfo{pages}{642} (\bibinfo{year}{1971}).

\bibitem[{\citenamefont{Blackley et~al.}(2014)\citenamefont{Blackley, Julienne,
  and Hutson}}]{Blackley:eff-range:2014}
\bibinfo{author}{\bibfnamefont{C.~L.} \bibnamefont{Blackley}},
  \bibinfo{author}{\bibfnamefont{P.~S.} \bibnamefont{Julienne}},
  \bibnamefont{and} \bibinfo{author}{\bibfnamefont{J.~M.}
  \bibnamefont{Hutson}}, \bibinfo{journal}{Phys. Rev. A}
  \textbf{\bibinfo{volume}{89}}, \bibinfo{pages}{042701}
  (\bibinfo{year}{2014}).

\bibitem[{\citenamefont{Gao}(1998)}]{Gao:QDT:1998}
\bibinfo{author}{\bibfnamefont{B.}~\bibnamefont{Gao}}, \bibinfo{journal}{Phys.
  Rev. A} \textbf{\bibinfo{volume}{58}}, \bibinfo{pages}{4222}
  (\bibinfo{year}{1998}).

\bibitem[{\citenamefont{Efimov}(1979)}]{Efimov1979lep}
\bibinfo{author}{\bibfnamefont{V.}~\bibnamefont{Efimov}},
  \bibinfo{journal}{Sov. J. Nuc. Phys.} \textbf{\bibinfo{volume}{29}},
  \bibinfo{pages}{546} (\bibinfo{year}{1979}).

\bibitem[{\citenamefont{Braaten et~al.}(2008)\citenamefont{Braaten, Hammer,
  Kang, and Platter}}]{Braaten2008tbr}
\bibinfo{author}{\bibfnamefont{E.}~\bibnamefont{Braaten}},
  \bibinfo{author}{\bibfnamefont{H.-W.} \bibnamefont{Hammer}},
  \bibinfo{author}{\bibfnamefont{D.}~\bibnamefont{Kang}}, \bibnamefont{and}
  \bibinfo{author}{\bibfnamefont{L.}~\bibnamefont{Platter}},
  \bibinfo{journal}{Phys. Rev. A} \textbf{\bibinfo{volume}{78}},
  \bibinfo{pages}{043605} (\bibinfo{year}{2008}).

\bibitem[{Pet()}]{PetrovLesHouches2010}
\bibinfo{note}{D.~S.~Petrov, in {\it Proceedings of the Les Houches Summer
  Schools, Session 94}, edited by C. Salomon, G. V. Shlyapnikov, and L. F.
  Cugliandolo (Oxford University Press, Oxford, England, 2013), e-print
  arXiv:1206.5752.}

\bibitem[{not({\natexlab{b}})}]{note:unstable}
\bibinfo{note}{Fits with both $T$ and $\lambda$ as free parameters are
  numerically unstable because the parameters are too strongly correlated.}

\bibitem[{not({\natexlab{c}})}]{note:weights}
\bibinfo{note}{We use weights inversely proportional to the square of the
  errors of the individual fits. This strongly favors the results obtained from
  the linear fits (first and third data points in Fig.~\ref{fig.kappas}). This
  is reasonable because these fits are more sensitive to the resonance peaks
  than the logarithmic fits, which are more sensitive to the wings of the
  resonance}.

\bibitem[{not({\natexlab{d}})}]{note:residualT}
\bibinfo{note}{In the original analysis, a zero-temperature model was applied
  to fit only a subset of data in the wings of the resonance, where temperature
  limitations remain small. This eliminated a large contribution to the
  temperature-induced shift.}

\bibitem[{\citenamefont{Julienne and Hutson}(2014)}]{Julienne:Li67:2014}
\bibinfo{author}{\bibfnamefont{P.~S.} \bibnamefont{Julienne}} \bibnamefont{and}
  \bibinfo{author}{\bibfnamefont{J.~M.} \bibnamefont{Hutson}},
  \bibinfo{journal}{Phys. Rev. A} \textbf{\bibinfo{volume}{89}},
  \bibinfo{pages}{052715} (\bibinfo{year}{2014}).

\bibitem[{\citenamefont{Le~Roy}(1998)}]{LeRoy:1998}
\bibinfo{author}{\bibfnamefont{R.~J.} \bibnamefont{Le~Roy}},
  \bibinfo{journal}{J. Mol. Spect.} \textbf{\bibinfo{volume}{191}},
  \bibinfo{pages}{223} (\bibinfo{year}{1998}).

\bibitem[{not({\natexlab{e}})}]{note:aeff}
\bibinfo{note}{The approximation of an effective scattering length from Ref.
  \cite{Wenz2009uti} leads instead to $a_-^*/r_{\rm vdW} = - 8.15(7)$, which is
  not consistent with our result based on the generalized STM equations. This
  shows the limited usefulness of the effective scattering length at the
  precision level of the present work.}

\bibitem[{\citenamefont{Th{\o}gersen
  et~al.}(2008{\natexlab{b}})\citenamefont{Th{\o}gersen, Fedorov, and
  Jensen}}]{Thogersen2008upo}
\bibinfo{author}{\bibfnamefont{M.}~\bibnamefont{Th{\o}gersen}},
  \bibinfo{author}{\bibfnamefont{D.~V.} \bibnamefont{Fedorov}},
  \bibnamefont{and} \bibinfo{author}{\bibfnamefont{A.~S.}
  \bibnamefont{Jensen}}, \bibinfo{journal}{Phys. Rev. A}
  \textbf{\bibinfo{volume}{78}}, \bibinfo{pages}{020501(R)}
  (\bibinfo{year}{2008}{\natexlab{b}}).

\bibitem[{\citenamefont{Axilrod and Teller}(1943)}]{Axilrod1943iot}
\bibinfo{author}{\bibfnamefont{B.~M.} \bibnamefont{Axilrod}} \bibnamefont{and}
  \bibinfo{author}{\bibfnamefont{E.}~\bibnamefont{Teller}},
  \bibinfo{journal}{J. Chem. Phys.} \textbf{\bibinfo{volume}{11}},
  \bibinfo{pages}{299} (\bibinfo{year}{1943}).

\bibitem[{\citenamefont{Sold\'{a}n et~al.}(2003)\citenamefont{Sold\'{a}n,
  Cvita\v{s}, and Hutson}}]{Soldan2003tbn}
\bibinfo{author}{\bibfnamefont{P.}~\bibnamefont{Sold\'{a}n}},
  \bibinfo{author}{\bibfnamefont{M.~T.} \bibnamefont{Cvita\v{s}}},
  \bibnamefont{and} \bibinfo{author}{\bibfnamefont{J.~M.}
  \bibnamefont{Hutson}}, \bibinfo{journal}{Phys. Rev. A}
  \textbf{\bibinfo{volume}{67}}, \bibinfo{pages}{054702}
  (\bibinfo{year}{2003}).

\bibitem[{\citenamefont{D'Incao et~al.}(2009)\citenamefont{D'Incao, Greene, and
  Esry}}]{Dincao2009tsr}
\bibinfo{author}{\bibfnamefont{J.~P.} \bibnamefont{D'Incao}},
  \bibinfo{author}{\bibfnamefont{C.~H.} \bibnamefont{Greene}},
  \bibnamefont{and} \bibinfo{author}{\bibfnamefont{B.~D.} \bibnamefont{Esry}},
  \bibinfo{journal}{J. Phys. B} \textbf{\bibinfo{volume}{42}},
  \bibinfo{pages}{044016} (\bibinfo{year}{2009}).

\bibitem[{\citenamefont{Dyke et~al.}(2013)\citenamefont{Dyke, Pollack, and
  Hulet}}]{Dyke2013frc}
\bibinfo{author}{\bibfnamefont{P.}~\bibnamefont{Dyke}},
  \bibinfo{author}{\bibfnamefont{S.~E.} \bibnamefont{Pollack}},
  \bibnamefont{and} \bibinfo{author}{\bibfnamefont{R.~G.} \bibnamefont{Hulet}},
  \bibinfo{journal}{Phys. Rev. A} \textbf{\bibinfo{volume}{88}},
  \bibinfo{pages}{023625} (\bibinfo{year}{2013}).

\bibitem[{\citenamefont{Fletcher et~al.}(2013)\citenamefont{Fletcher, Gaunt,
  Navon, Smith, and Hadzibabic}}]{Fletcher2013soa}
\bibinfo{author}{\bibfnamefont{R.~J.} \bibnamefont{Fletcher}},
  \bibinfo{author}{\bibfnamefont{A.~L.} \bibnamefont{Gaunt}},
  \bibinfo{author}{\bibfnamefont{N.}~\bibnamefont{Navon}},
  \bibinfo{author}{\bibfnamefont{R.~P.} \bibnamefont{Smith}}, \bibnamefont{and}
  \bibinfo{author}{\bibfnamefont{Z.}~\bibnamefont{Hadzibabic}},
  \bibinfo{journal}{Phys. Rev. Lett.} \textbf{\bibinfo{volume}{111}},
  \bibinfo{pages}{125303} (\bibinfo{year}{2013}).

\end{thebibliography}

\end{document}